\documentclass[fleqn,usenatbib]{mnras}

\usepackage[T1]{fontenc}
\usepackage{ae,aecompl}

\usepackage{graphicx}	
\usepackage{amsmath}	
\usepackage{amssymb}	

\title[Impact of ISM structure on the slope of radio $\Sigma-D$ relation]{Interstellar medium structure and the slope of the radio $\Sigma-D$ relation of supernova remnants}

\author[P. Kosti\'c, B. Vukoti\'c, D. Uro\v sevi\'c, B. Arbutina and T. Prodanovi\'c]{
P. Kosti\'c,$^{1}$\thanks{E-mail:
	perakostic@gmail.com (PK), bvukotic@aob.rs (BV), dejanu@matf.bg.ac.rs (DU), arbo@matf.bg.ac.rs (BA), prodanvc@df.uns.ac.rs (TP).}
B. Vukoti\'c,$^{2}$
D. Uro\v sevi\'c,$^{1,4}$
B. Arbutina$^{1}$
and T. Prodanovi\'c$^{3}$
\\
$^{1}$ Department of Astronomy, Faculty of Mathematics, University of Belgrade, Studentski trg 16, 11000 Belgrade, Serbia\\
$^{2}$ Astronomical Observatory, Volgina 7, 11060 Belgrade 38, Serbia\\ 
$^{3}$ Department of Physics, Faculty of Sciences, University of Novi Sad,
Trg Dositeja Obradovica 4, 21000 Novi Sad, Serbia\\
$^{4}$ Isaac Newton Institute of Chile, Yugoslavia Branch
}

\date{Accepted 2016 June 06. Received 2016 June 03; in original form 2016 March 07}

\pubyear{2016}

\begin{document}
\label{firstpage}
\pagerange{\pageref{firstpage}--\pageref{lastpage}}
\maketitle

\begin{abstract}
We analyze the influence of fractal structure of the interstellar matter (ISM) density on the parameter values for the radio surface brightness to diameter ($\Sigma-D$) relation for supernovae remnants (SNRs). We model a dense ISM as a molecular cloud with fractal density structure. SNRs are modelled as spheres of different radius scattered in the modelled ISM. The surface brightness of the SNRs is calculated from the simple relation $\Sigma \propto \rho^{0.5}D^{-3.5}$ and also from the parametrized more general form $\Sigma \propto \rho^{\eta}D^{-\beta_0}$. Our results demonstrate that  empirical $\Sigma-D$ slopes that are steeper than the ones derived from theory, might be partly explained with the fractal structure of the ambient medium into which SNRs expand. The slope of the $\Sigma-D$ relation steepens if the density of the regions where SNRs are formed is higher. The simple geometrical effects combined with the fractal structure of the ISM can contribute to a steeper empirical $\Sigma-D$ slopes, especially for older remnants, and this is more pronounced if $\Sigma$ has a stronger dependence on ambient density.
\end{abstract}




\section{Introduction}
\label{Introduction}

There are various methods for determining of distance to supernovae remnants (SNRs). One of them is a statistical method  that calibrates the dependence between the radio surface brightness ($\Sigma$) and the diameter ($D$) of the remnant, called  the $\Sigma-D$ relation \citep[derived by][]{1960AZh....37..256S}. This method is often used for determining distances to supernova remnants in the Milky Way and other galaxies \citep{CaseBhattacharya1998,Arbutina2004,ArbutinaUrosevic2005,Urosevic2005,Urosevic2010,Vukotic2014,Pavlovicetal2013,Pavlovicetal2014}. For a given radio frequency ($\nu$) the dependence between the radio surface brightness ($\Sigma$) of an SNR and its diameter can be written in a power-law form as:
\begin{equation}
	\Sigma _\nu( D ) = A{D^{ - \beta }}, \label{gendef}
\end{equation}
where $\Sigma _\nu$ is in ${\rm{W}}\,{{\rm{m}}^{ - 2}}\,{\rm{H}}{{\rm{z}}^{ - 1}}\,{\rm{s}}{{\rm{r}}^{ - 1}}$, $D$ is in pc, while $A$ and $\beta$ are parameters determined either from observations or theoretical models.

The surface brightness does not depend on the SNR distance and can be calculated from the quantities that are directly measured with radio observations
as ${\Sigma _\nu } \propto {S_\nu } / {{\theta ^2}}$ where $S_\nu$ is the flux density and $\theta$ is the the SNR angular diameter. Therefore, from the observed $\Sigma$, the value of $D$ can be determined through the $\Sigma-D$ relation and  the distance to a SNR calculated as $d \simeq D/\theta$. Although, at first, some five decades ago, the $\Sigma-D$ relation was seen as a promising tool for distance determination, after many observed SNRs in our and other galaxies, it was clear that empirical $\Sigma-D$ relation was subject to a severe data scatter \citep{Green1984}. In addition, there is a significant discrepancy between $A$ and $\beta$ values from empirical and theoretical relations. Nevertheless, the consistent $\Sigma-D$ relation would present an indispensable tool  for the SNR distance determination especially in the Galaxy. On the other hand, studying the $\Sigma-D$ relation is important for understanding the evolution of synchrotron radiation of SNRs and related phenomena occurring at collisionless shock waves \citep{Bell1978,BerezkoVolk2004}.  

Relation (\ref{gendef}) is commonly expressed in log-log scale as
\begin{equation}
	\log \Sigma  = \log A - \beta \log D, \label{fitdef}
\end{equation}
because both axes extend over several orders of magnitude. In this way, $\beta$ represents the slope in the $\log \Sigma-\log D$ diagram, with intercept being $\log A$.

The relation given by \citet{1960AZh....37..256S}, for a synchrotron emission of a spherically expanding shell nebulae, is $\Sigma  \propto {D^{-6}}$ for a spectral index $\alpha=0.5$ ($S_{\nu} \propto \nu^{-\alpha}$). The number of different theoretical and empirical relations give the value of $\beta$ in the range of $ \sim 2$ to $ \sim 6$ \citep{PovedaWoltjer1968,ReynoldsChevalier1981,DuricSikvist1986,CaseBhattacharya1998,Guseinovetal2003,BerezkoVolk2004,Pavlovicetal2013,Pavlovicetal2014}. The evolution of the surface brightness and the diameter of a supernova remnant depends on various factors: total kinetic energy (and type) of supernova, ejected mass, ambient density, geometry of magnetic field, injection and distribution of relativistic electrons, etc. These factors are different for different supernovae (SNe) and so a large range of $\beta$ (and also $A$) is to be expected. 

According to \citet{ArbutinaUrosevic2005}, an ideal case where a SNR expands into a constant density ISM, should result in a roughly parallel evolutionary $\Sigma-D$  tracks, where remnants expanding into a denser interstellar matter might have stronger radio synchrotron emission and consequently a larger $\Sigma$. Considering that molecular clouds are the places of star formation, we can assume that massive stars, which have short lives, explode as supernovae in (or near) the dense medium of molecular clouds. All theoretical $\Sigma-D$ relations were derived for a homogeneous ambient density, but generally, the density of the SNR environment is not homogeneous on a typical SNR size scale, especially in a dense interstellar matter regions typical of insides of molecular clouds. There, the ISM gas is in the state of constant turbulent motions which create wide (power-law) spectrum of velocities and densities at different length scales. All these properties characterize the highly inhomogeneous fractal structure which was observed and studied for several decades \citep{BazellDesert1988,Scalo1990,Vogelaaretal1991,Beech1992,ElmegreenFalgarone1996,Elmegreen1997,Stutzki1998,Sanchez2005,Sanchez2006,Sanchez2007,Federrathetal2009,Seon2012}. The main aim of this paper is to extend on that and to show that such fractal environment can affect the slope $\beta$ of the $\Sigma-D$ relation. Here we present a toy $\Sigma-D$ model for a shell type SNRs expanding in a medium of fractal structure and density according to Elmegreen's model of molecular clouds.

We argue that implications of our simplistic approach suggest that fractal nature of the ISM should not be ignored in the $\Sigma-D$ evolution studies of SNRs because it can result in a significant change of the $\Sigma-D$ slope depending on the structure of the ambient ISM.

\section{$\Sigma-D$ relations for supernova remnants}

The $\Sigma-D$ parameter $A$ depends on the SN features like explosion energy, ejected mass, SN type, etc. and the ambient ISM features like density, intensity (and shape) of magnetic field, etc. The common assumption is that these features hardly affect the coefficient $\beta$ (which will be the main topic of this paper), as $\beta$ is mainly affected by the magnetic field evolution in SNR and particle acceleration processes. However, if $\Sigma$ depends on the ambient ISM density $\rho$, then a potential $\rho(D)$ dependence would imply an additional $\Sigma(D)$ dependence on top of the theoretically derived $\Sigma(D)$. Indeed, as we will demonstrate in this paper, the fractal density structure of the ISM results in $\rho(D)$ dependence that can in turn, for the assumed $\Sigma(\rho)$ dependence, significantly change the slope $\beta$ compared to the cases when $\rho$ is considered to be constant.

It is clear that there is no universal $\Sigma-D$ relation, but different ones for different kinds of SNRs and their ambient conditions can exist. \citet{ArbutinaUrosevic2005} found two different $\Sigma-D$ relations for SNRs associated with low and high ambient densities. The relation for the higher densities is shifted towards higher surface brightness, which is expected due to the relation $\Sigma  \propto {\rho ^\eta },\;({\eta  \ge 0} )$, mentioned above. For high densities\footnote{All theoretical $\Sigma-D$ relations are given assuming $\alpha=0.5$.} they obtained
	\begin{equation}
		\Sigma _{1\;{\rm{GHz}}} = 2.2_{ - 1.3}^{ + 3.1} \times {10^{ - 15}}{D^{ - 3.3 \pm 0.4}}\;{\rm{W}}\,{{\rm{m}}^{ - 2}}\,{\rm{H}}{{\rm{z}}^{ - 1}}\,{\rm{s}}{{\rm{r}}^{ - 1}},
	\end{equation}
	while for low densities,
	\begin{equation}
		\Sigma _{1\;{\rm{GHz}}} = 3.9_{ - 2.9}^{ + 11.3} \times {10^{ - 17}}{D^{ - 3.2 \pm 0.6}}\;{\rm{W}}\,{{\rm{m}}^{ - 2}}\,{\rm{H}}{{\rm{z}}^{ - 1}}\,{\rm{s}}{{\rm{r}}^{ - 1}}.
	\end{equation}
	
	Theoretical $\Sigma-D$ relation derived by \citet{BerezkoVolk2004} has a slope of $\beta  = 4.25$ in early Sedov phase and is independent of the ambient density. However, they also note that due to a shock modification there is a small density dependence for ambient densities above ${N_{\rm{H}}} = 0.003\;{\rm{cm}}^{-3}$ for which the slope is slightly steeper. Similar dependence exists in $\Sigma-D$ relations simulated in this paper as a consequence of hierarchical distribution of matter in a fractal cloud. The indication of such a dependence is also seen in empirical relations from \citet{Pavlovicetal2013} for 28 SNRs in molecular clouds associated with dense ambient medium, 
	\begin{equation}
		\Sigma _{1\;{\rm{GHz}}} = 3.89_{ - 2.98}^{ + 12.81} \times {10^{ - 15}}{D^{ - 3.9 \pm 0.4}}\;{\rm{W}}\,{{\rm{m}}^{ - 2}}\,{\rm{H}}{{\rm{z}}^{ - 1}}\,{\rm{s}}{{\rm{r}}^{ - 1}}, \label{Pavlovic}
	\end{equation}
	and for 5 Balmer dominated SNRs associated with rarefied ambient medium,
	\begin{equation}
		\Sigma _{1\;{\rm{GHz}}} = 1.89_{ - 1.29}^{ + 4.08} \times {10^{ - 16}}{D^{ - 3.5 \pm 0.5}}\;{\rm{W}}\,{{\rm{m}}^{ - 2}}\,{\rm{H}}{{\rm{z}}^{ - 1}}\,{\rm{s}}{{\rm{r}}^{ - 1}}.
	\end{equation}
	From these it can be seen that both, coefficient $A$ and $\beta$, can depend on ambient density.
	In our work we follow theoretical results of \citet{DuricSikvist1986} and express the radio surface brightness as a function of ambient density $\rho$ and $D$, which, for a typical shell type remnant in the adiabatic (Sedov) phase, has the form:
	
	\begin{equation}
		\Sigma _{1\,{\rm{GHz}}}\left( {\rho ,D} \right) = 4 \times {10^{ - 15}}{\left( {\frac{\rho }{{{\rho _0}}}} \right)^{0.5}}{D^{ - 3.5}},\;\left( {D \gg 1\;{\rm{pc}}} \right), \label{sig}
	\end{equation}
	where $\Sigma_{1\,{\rm{GHz}}}$ is in ${\rm{W}}\,{{\rm{m}}^{ - 2}}\,{\rm{H}}{{\rm{z}}^{ - 1}}\,{\rm{s}}{{\rm{r}}^{ - 1}}$, $D$ is in pc, and $\rho_0 = 10^{-24}$ g cm$^{-3}$.
	The Equation (\ref{sig}) will be used for calculating the surface brightness in our simulations. Due to a number of theoretical and empirical $\Sigma-D$ relations that can be found in literature we also utilize:
	\begin{equation}
		\Sigma \propto D^{-\beta}\propto \rho^\eta D^{-\beta_0},
	\end{equation} 
	as a more general form of Equation (\ref{sig}).
	
From this it is evident that various factors can influence the parameter $\beta$ from Equation (\ref{gendef}). In addition, the estimated values of $\beta$ may differ significantly depending on the applied fitting procedure used in Eq. (\ref{fitdef}) \citep[see][]{Pavlovicetal2013}. Usually, the regression is performed on the data for which it is assumed that it can be represented by functional dependence of the form $Y = f(X)$, where $X$ is the independent variable determined with a high precision, and even more important, it does not depend on some other parameters of the examined phenomena. Normally, one would then perform the regression such as to minimize data offsets from the fit line along the $Y$-axis, usually termed vertical regression. However, the $\Sigma-D$ relation should not be treated as having a functional dependence in the form $\Sigma = f (D)$ or $D = f (\Sigma)$ because none of the variables can be considered as independent. Both, $\Sigma$ and $D$, are SNR parameters that can depend on various SNR properties (discussed earlier in this section), and thus cannot be estimated without significant uncertainties. A better choice would be to treat both variables symmetrically such that a regression offset of the data point from the fit line is calculated along the direction perpendicular to the fit line and not just along the $Y$-axis. This is called the orthogonal regression. The results from \citet{Pavlovicetal2013} showed that the orthogonal regression gives the most stable fit parameters estimates and should be favoured over the other types of regressions.
	
On the other hand, the $\Sigma-D$ relation is most often (almost in all cases) calibrated with a vertical regression. However this type of regression tends to underestimate the slope $\beta$ especially at its larger values \citep{Pavlovicetal2013}. Thus, we select the orthogonal regression as a preferred fitting procedure for the work done in this paper.
	
We model the SNRs as spheres of different sizes scattered in the ISM model of fractal density structure. The surface brightness of the modelled SNRs is then determined from the ambient density values of the regions subtended by the SNRs. The  $\Sigma-D$ relations are then obtained using Eq. (\ref{sig}). 

In Section \ref{ISM model} we present the properties of the fractal medium, the fractal molecular cloud model and technical details of the simulation. The results are given and  discussed in Section \ref{Results}. Other discussions concerning the presented model and some implications of the results will take place in Section \ref{Discussion} and conclusions of this work are given in the last section.

\section{Model}
\label{ISM model}

In our analysis we use the model of the fractal molecular cloud given by \citet{Elmegreen1997}. However, first some important properties of interstellar clouds will be discussed.

\subsection{Properties of ISM clouds}
\citet{ElmegreenFalgarone1996} gave a short review of interstellar matter structure noting that the standard cloud model (separate and distinct clouds - \textit{clumps} - embedded in a warm tenuous intercloud medium) is oversimplified and that decades of observations changed our perception of the interstellar medium \citep[see also][]{Scalo1990}. They pointed out that clouds are neither uniform nor isolated and their boundaries are usually convoluted and fractal. The clouds are hierarchically structured, i.e. contain self-similar substructures on several levels of hierarchy without clear boundaries. The self-similarity seems to be the crucial property of interstellar medium because the distributions of mass and size of the clouds and their fragments appear as power laws which are known to be scale-free. Numerous authors have studied the structure of molecular clouds, finding the self-similar scaling and other independent evidences of their clumpy structure \citep[and the references therein]{BazellDesert1988,Stutzkietal1988,Howeetal1991,Scalo1990,Dickmanetal1990,Falgaroneetal1991,Falgarone1992,ElmegreenFalgarone1996,Cubicketal2008}. The fractal structure and geometry of the ISM arises from the action of interstellar turbulence \citep{Stutzki1998,OssenkopfMacLow2002,HeyerBrunt2004,Federrathetal2009,Federrathetal2010,Roman-Duval2011,Federrath2013}. The physical origin of the turbulent fluctuations is still under discussion, but it is known that physical conditions in the ISM, especially the compressibility of the gas, alter the statistics established for the incompressible Kolmogorov turbulence \citep{Federrathetal2009}. \cite{Federrathetal2009} contributed to the understanding of different and extreme ways of driving the interstellar turbulence. Their numerical experiments showed that the driving mode of the turbulence affects the fractal properties of the molecular clouds.

Power law distributions of mass and size in the molecular clouds exist only if the dynamics is dominantly governed by turbulence and not by self-gravity or magnetic fields (e.g. in the case of extremely dense cores that will form protostars). Therefore, there is a large range of masses and sizes at which clouds are considered fractal. The size distribution is given by
\begin{equation}
	n\left( S \right)dS = {S^{ - (1 + {D_{\rm{f}}})}}dS, \label{nS}
\end{equation}
applies on scales $0.01 - 100\;{\rm{pc}}$, while the mass distribution
\begin{equation}
	n\left( M \right)dM = {M^{ - (1 + {{{D_{\rm{f}}}} \mathord{\left/{\vphantom {{{D_{\rm{f}}}} \kappa }} \right.\kern-\nulldelimiterspace} \kappa })}}dM \label{nM}
\end{equation}
applies for masses in the range ${10^{ - 2}} - {10^7}\;{M_ \odot }$ \citep{Elmegreen1997}. Evidently, the structure sizes cover almost all scales - from giant molecular clouds (GMCs), down to their tiniest fragments. In Eqs. (\ref{nS}) and (\ref{nM}) $S$ and $M$ are the size (diameter) and mass of the cloud (or a clump), $D_{\rm{f}}$ is a fractal dimension (explained in the next paragraph) and $\kappa$ is the exponent of the mass-size relation
\begin{equation}
	M \propto {S^\kappa }. \label{M-S}
\end{equation}

The self-similarity of a fractal object is quantitatively described by its fractal (or similarity) dimension. This, generally non-integer number, characterizes the scaling ratio between a structure and its substructures, and the object's space-filling ability. Let us make an analogy with Euclidean dimension. If we have a 3-dimension object like a cube, we can split it into $N$ subcubes with $L$ times smaller edge (size) such that
\begin{equation}
N = L^3.
\end{equation}		
Now, in a fractal with a fractal dimension $D_{\rm{f}}$, every structure contains $N$ self-similar substructures of $L$ times smaller size such that
\begin{equation}
N = L^{D_{\rm{f}}}. \label{N-L}
\end{equation}
In the studies of fractal structure of molecular clouds, the quantity $D_{\rm{f}}$ is usually called the \textit{volume} fractal dimension (as a measure of the volume-filling ability) and the \textit{box-counting} dimension or ''capacity`` (by the box-coverage method of its estimation). This topic is nicely explained in \cite{Voss1988}. However, this is not the only definition of the fractal dimension. In general, any fractal quantity $A$ can have an associated dimension $D_A$ so it relates to the size scale $S$ as
\begin{equation}
A \propto S^{D_A}. \label{A-S}
\end{equation}
For example, in the case of describing a mass-size relation as in Equation (\ref{M-S}), we have a mass dimension $\kappa \equiv D_M$. 

One of the main methods of measuring the fractal dimension of the ISM so far was the perimeter-area relation of the iso-brightness contours (boundaries) of clouds images \citep[see][]{BazellDesert1988,Dickmanetal1990,VogelaarWakker1994}. This \textit{perimeter-based} dimension characterizes the irregularity of the boundary. We will explain it here briefly, following the approach of \citet{Falgaroneetal1991}, which originates from \citet{Lovejoy1982} and \citet{Mandelbrot1977}. The relation between the perimeter $P$ and the area $A$ of classical (smooth) planar figures, such as a circle or a square, is $P \propto A^{1/2}$. By definition, for a planar curve with the fractal dimension $D_{\rm{per}}$ this relation is $P \propto A^{{D_{\rm{per}}} /2}$, so for a circle or a square $D_{\rm{per}} = 1$ while for the extremely convoluted curve it would tend towards 2 ($P \propto A$ when the curve completely fills the plane). The same applies for the surface in 3-dimensional space, the surface of the object relates to its volume as $S \propto V^{{D_{\rm{sur}}} /3}$, so for a sphere or a cube $D_{\rm{sur}} = 2$, tending towards 3 for convoluted surface. It is known that plane intersections of a surface of dimension $D_{\rm{sur}}>2$ give contours of dimension $D_{\rm{per}} = D_{\rm{sur}} - 1$, but the same relation cannot be claimed for plane projections of a surface (which is the case with images of clouds).

In the literature, the measured values for the perimeter-based fractal dimension of the ISM are within interval $D_{\rm{per}} \sim 1.2-1.6$ over a wide range of scales mentioned above, and at different distances. However, most of them are $D_{\rm{per}} \approx 1.3-1.4$. Many authors \citep[e.g.][]{ElmegreenFalgarone1996} used the simple (but unreliable) relation $D_{\rm{f}} = D_{\rm{per}} + 1$ to obtain the volume fractal dimension (notice that we don't really know if $D_{\rm{f}}$ and $D_{\rm{sur}}$ are necessarily of the same value). Thus, for $D_{\rm{per}} = 1.3$ it is assumed that the volume fractal dimension is ${D_{\rm{f}}} = 2.3$. However, in a very detailed analysis and using numerical simulations, \cite{Sanchez2005} explain the difference and relationship between various types of fractal dimensions. Their results show that these two dimensions are not related in such a way and that $D_{\rm{per}}$ even decreases with the increase of $D_{\rm{f}}$, so for $D_{\rm{per}} = 1.35$ they get ${D_{\rm{f}}} = 2.6$. The results of \cite{Sanchez2007} support the relatively high average fractal dimension for the ISM, $2.6 \leq D_{\rm{f}} \leq 2.8$. In high-resolution numerical experiments of interstellar turbulence, \cite{Federrathetal2009} obtain two values for the (box-counting) fractal dimension, depending on the driving mode of the turbulence: ${D_{\rm{f}}} \approx 2.6$ for (usually adopted) solenoidal (divergence-free) forcing, and ${D_{\rm{f}}} \approx 2.3$ for compressive (curl-free) forcing. This means that solenoidal driving leads to more volume-filling structures, while compressive driving results in more sheet-like structures with larger voids. They also calculated projected perimeter-area dimensions for these two driving modes and the values show similar dependence: ${D_{\rm{per}}} \approx 1.36$ for solenoidal forcing, and ${D_{\rm{per}}} \approx 1.18$ for compressive forcing \citep[notice the negative correlation with results from][]{Sanchez2005}.

\citet{ElmegreenFalgarone1996} gave the values of exponents $\kappa$ and $1 + {{{D_{\rm{f}}}} \mathord{\left/{\vphantom {{{D_{\rm{f}}}} \kappa }} \right.\kern-\nulldelimiterspace} \kappa } = \alpha_{\rm{M}}$, and fractal dimension $D_{\rm{f}}$, determined from numerous cloud surveys in the literature. The exponents were found to be roughly in the range $2.2<\kappa<2.5$ (for the all-cloud plot, but for some surveys it goes up to 3.7), $1.6<{\alpha_M}<2$ and $D_{\rm{f}}=2.3\pm0.3$. The authors also noted that for self-similar fractals $\kappa (\equiv D_M)$ is equal to $D_{\rm{f}}$ which will be the case in our model of the cloud. Following the results of \cite{Federrathetal2009}, we will run the simulations for both $D_{\rm{f}}=2.3$ and $D_{\rm{f}}=2.6$. Other parameters of the cloud model, listed in Table \ref{table:1}, are explained in the following section.

\subsection{Fractal cloud model}
\label{model}
In this section we present a model of the molecular cloud which has a fractal-like density structure. The recipe for modelling the fractal cloud is adopted from \citet{Elmegreen1997} and partly from \citet{Sanchez2006}. In order to make a hierarchically structured cloud, for each $i$ of $H$ levels of hierarchy we randomly place $N$ points inside a box of size $1\times1\times1$. The positions of points for $i=1$ on \textit{x}-axis are ${x_1} = 0.5 + 2\left( {{r_1} - 0.5} \right) / \mathord{L}$, so the second level ($i=2$) positions are ${x_2} = {{x_1} + 2\left( {{r_2} - 0.5} \right)} \mathord{/ {{L^2}}}$, the third ${x_3} = {{x_2} + 2\left( {{r_3} - 0.5} \right)} \mathord{/ {{L^3}}}$, up to a final hierarchy level $i=H$, ${x_H} = {{x_{H - 1}} + 2\left( {{r_H} - 0.5} \right)} / \mathord{{{L^H}}}$, where for $i = {1,...,H}$, ${r_{i}}$ are the random positions within $[0,1]$ interval for every single point, and $L>1$ is a geometric factor for reducing the size of the next level (let us call it the scale factor). In other words, first we place $N$ points in the box, then every of these points becomes the centre of the $L$ times smaller sub-box where we place another $N$ points, and so on, until we reach the last level. This way, the total number of points at the final level is $N^H$, and these are the only points that are left in the box at the end. The same procedure is done for the \textit{y}- and \textit{z}-axis. The volume fractal dimension of this structure is given as ${D_{\rm{f}}} = {{\log N} / {\log L}}$ (see Eq. \ref{N-L}). For $N=12$ and $L=3$ its value is ${D_{\rm{f}}} \approx 2.3$ ($L=2.6$ gives ${D_{\rm{f}}} = 2.6$). The number of levels $H$ determine the maximum density contrast and for the interstellar matter its value is within interval $H \sim 3 - 6$. We take $H=4$ (see Table \ref{table:1}).

\begin{figure}
	\centering
	\includegraphics[width=0.45\textwidth]{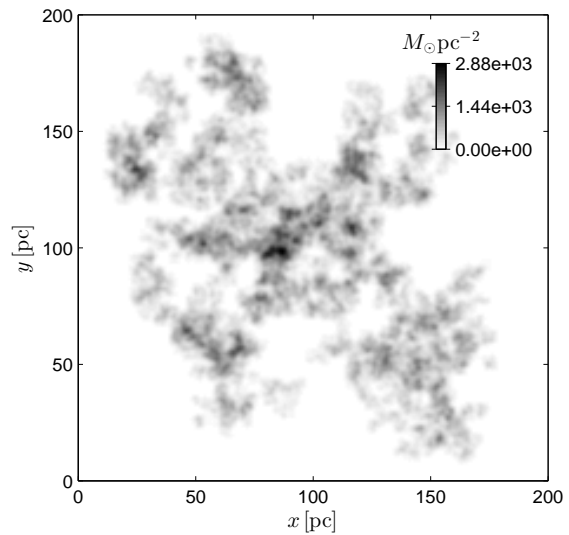}
	\caption{An example of a modelled fractal cloud of 200 {\rm{pc}} diameter ($D_{\rm{f}}=2.3$). It shows the projection of density (the column density) in a box at the \textit{z}-axis (line of sight).} \label{oblak}
\end{figure}

Now we created a massless cloud that consist of $N^H$ points with fractal distribution. As in \cite{Sanchez2006}, in order to obtain the mass density distribution inside the cloud, we convolve all the points with the Gaussian kernel function (with integral of 1) and calculate the density through the whole volume in the box. The density at the position (${x_i},{y_i},{z_i}$) is
\begin{equation}
	f\left( {{x_i},{y_i},{z_i}} \right) = \frac{1}{{{N^H}{\sigma ^3}{{\left( {2\pi } \right)}^{{3 \mathord{\left/				{\vphantom {3 2}} \right.} 2}}}}}
	\times \sum\limits_{j = 1}^{{N^H}} \exp \left\{{ { - \frac{1}{{2{\sigma ^2}}}r_{ij}^2} }\right\} \label{polje}
\end{equation}
where $\sigma$ is chosen to be about two orders of magnitude smaller than the box edge \citep[according to][]{Sanchez2006}, $\left( {{x_i},{y_i},{z_i}} \right)$ are the grid coordinates of the $i^{\rm{th}}$ cell and $r_{ij}^2 = {( x_i - x_j )}^2 + {( y_i - y_j )}^2 + {( z_i - z_j )}^2$. The box is scaled to the size of the cloud in parsecs and the cell size is 1 pc. Because the integrated density of the box is 1, by multiplying every cell's density with the mass of the cloud in $M_ \odot$ (which is calculated from the mass-size relation) we obtain the density of every cell in ${M_ \odot }\,{\rm{p}}{{\rm{c}}^{ - 3}}$. With this procedure we modelled the density scalar field of the molecular cloud in 3D,  the projection of which is shown in Figure \ref{oblak}. Relevant parameters and relations are given in Table \ref{table:1} and \ref{table:2}.

\begin{table}
	\begin{center}
		\caption{List of parameters of the cloud model. $D_{\rm{f}}$ - fractal dimension, $\kappa$ - mass-size relation exponent, $N$ - number of points at a level of hierarchy, $L$ - scale factor, $H$ - number of levels, $\sigma$ - smoothing factor. $S_{{\rm{cl}}}$ is the size of the cloud which corresponds to the edge of the box (see Section \ref{model}).}
		\label{table:1}
		\begin{tabular}{ |c|c|c| } 
			\hline
			Parameter & Value & Reference \\ 
			\hline
			$D_{\rm{f}}$ (compressive) & 2.3 & \cite{Federrathetal2009} \\ 
			$D_{\rm{f}}$ (solenoidal) & 2.6 & \cite{Federrathetal2009} \\
			$\alpha_M$ & 2 & from $\kappa = D_{\rm{f}}$ \\ 
			$N$ & 12 & \cite{Elmegreen1997} \\ 
			$L$ (compressive) & 3 & $=N^{1/D_{\rm{f}}}$ \\ 
			$L$ (solenoidal) & 2.6 & $=N^{1/D_{\rm{f}}}$ \\
			$H$ & 4 & \cite{Elmegreen1997} \\ 
			$\sigma$ & $0.005{S_{{\rm{cl}}}}$ & \cite{Sanchez2006} \\ 
			\hline
		\end{tabular}
	\end{center}
\end{table}

\begin{table}
	\begin{center}
		\caption{Distributions and relations for a modelled cloud. Respectively, these are: size distribution, mass distribution, mass-size relation and density-size relation. $R$ is the radius of the cloud and it is taken to be $R=S_{{\rm{cl}}}/2$. Mass-size relation is derived from the condition that a cloud of a radius 1 pc has an average number density of 1000 ${{\rm{H}}_2}\,{\rm{cm}}^{-3}$. Note: The numerical exponents in the right column are for $D_{\rm{f}}=2.3$, and the ones in parentheses are for $D_{\rm{f}}=2.6$.}
		\label{table:2}
		\begin{tabular}{ |c|c| } 
			\hline
			Distribution/Relation & Parametrization \\ 
			\hline
			$n\left( S \right)dS = {S^{ - (1 + {D_{\rm{f}}})}}dS$ & $n\left( R \right)dR = {R^{ - 3.3(-3.6)}}dR$ \\ 
			$n\left( M \right)dM = {M^{ - (1 + {{{D_{\rm{f}}}} \mathord{\left/{\vphantom {{{D_{\rm{f}}}} \kappa }} \right.\kern-\nulldelimiterspace} \kappa })}}dM$ & $n\left( M \right)dM = {M^{ - 2}}dM$ \\ 
			$M \propto {S^\kappa }$ & $\frac{M}{[M_ \odot]} = 210\left({\frac{R}{[\rm{pc}]}}\right)^{2.3(2.6)}$ \\ 
			$\rho  \propto {S^{\kappa  - 3}}$ & $\frac{\rho}{[M_ \odot \rm{pc}^{-3}]} = 50\left({\frac{R}{[\rm{pc}]}}\right)^{-0.7(-0.4)}$ \\ 
			\hline
		\end{tabular}
	\end{center}
\end{table}

\subsection{Simulations of supernova remnants}
In every modelled cloud we simulated the sample of 100 supernova remnants by placing spheres with centres randomly positioned inside the box volume, under the condition that they are in cells within a given density range (which we named \textit{central densities}, $\rho_\mathrm{C}$). The SNR radii are chosen randomly on the logarithmic scale between 5 and 50 pc, in order to obtain a uniform distribution of points on the $\log D$ axis of the $\Sigma-D$ diagram. The ambient density needed to calculate the surface brightness (using the Eq. \ref{sig}) was taken to be the average density inside the sphere. Three sizes of clouds (box edges $S_{\rm{cl}}$) are modelled: 50, 100 and 200 pc, and the central density ranges were divided in six logarithmic scale bins, from 1 to 1000  ${\rm{H}}_2\,{\rm{cm}}^{-3}$ or $\log \left( {{{{\rho _{\rm{C}}}} \mathord{\left/{\vphantom {{{\rho _{\rm{C}}}} {[{{\rm{H}}_2}\,{\rm{cm}}^{-3}]}}} \right.\kern-\nulldelimiterspace} {[{{\rm{H}}_2}\,{\rm{cm}}^{-3}]}}} \right) = [0.25 \pm 0.25,\;0.75 \pm 0.25,\;1.25 \pm 0.25,\;1.75 \pm 0.25,\;2.25 \pm 0.25,\;2.75 \pm 0.25]$. As the star formation mostly takes place in the giant molecular clouds, the cloud diameters are chosen accordingly in such a way to represent the large span of GMC sizes. For every cloud size and $\rho_{\rm{C}}$ we simulated 100 SNRs and by fitting their $\Sigma-D$ relation, values for $\log{A}$ and $\beta$ were obtained. To obtain robust results, free of statistical fluctuations, the values of $\log{A}$ and $\beta$ were calculated as medians of 100 such simulations, each one with newly simulated cloud. The error for $\log{A}$ and $\beta$ was taken so that it includes $70\%$ of obtained values around the median ($35\%$ from both sides of median in the sorted array of obtained values). All $\Sigma-D$ relations were fitted with orthogonal regression, although because of a relatively small scatter of points there is no significant difference between vertical and orthogonal fits (vertical gives a bit flatter slopes, but mostly by about one percent, $(\beta_{\rm{o}}-\beta_{\rm{v}})/\beta_{\rm{o}} < 0.015$, where $\beta_{\rm{o}}$ and $\beta_{\rm{v}}$ are the slopes from orthogonal and vertical regression, respectfully). The significance of orthogonal regression is more apparent when fitting real observational data, where the scatter is much larger. The whole procedure is done separately for the fractal dimension values of 2.3 and 2.6. The results are presented in Figure \ref{result} and Section \ref{Results}.

\section{Results}
\label{Results}

\begin{figure*}
	\centering
	\includegraphics[width=0.48\textwidth]{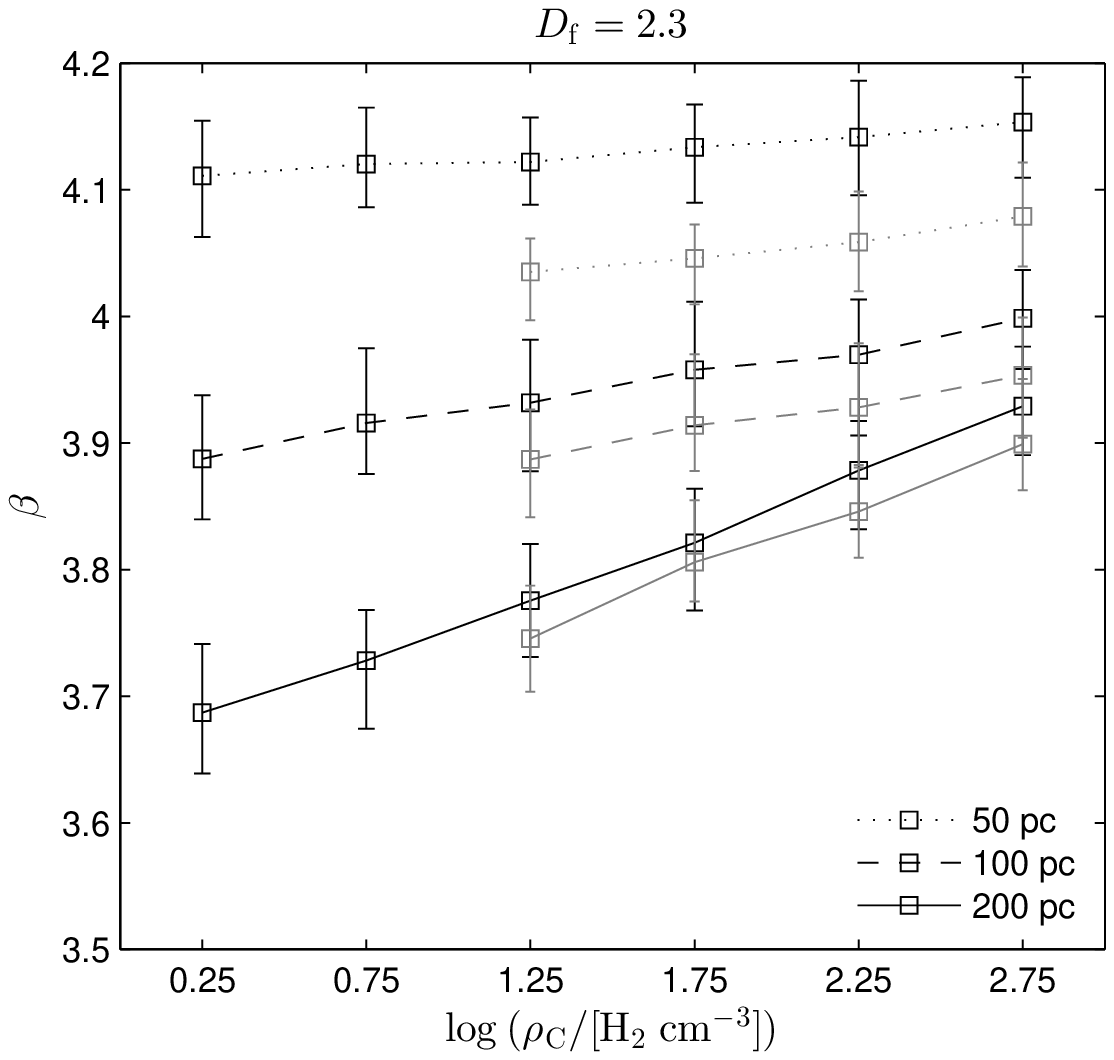}
	\includegraphics[width=0.48\textwidth]{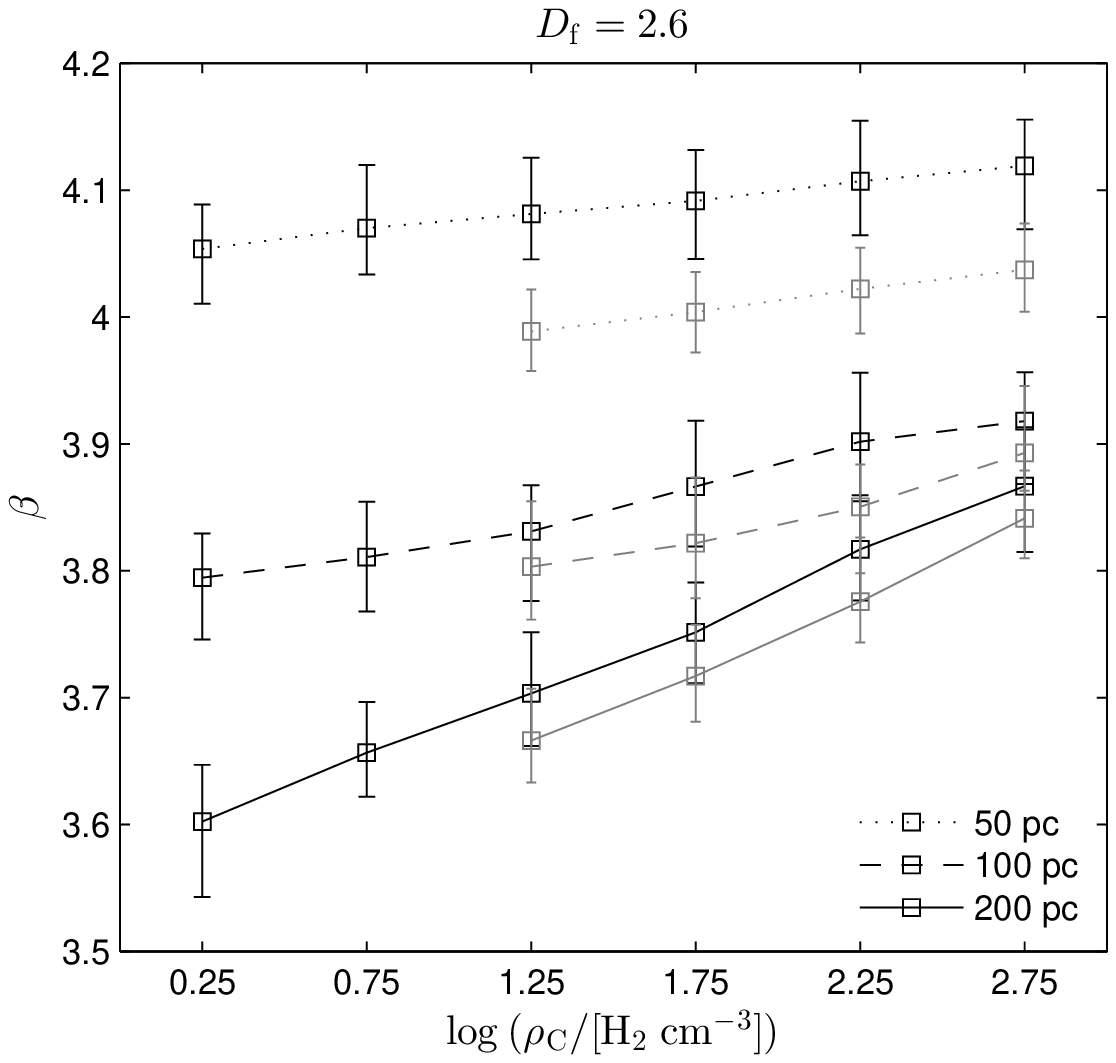} \\
	\includegraphics[width=0.48\textwidth]{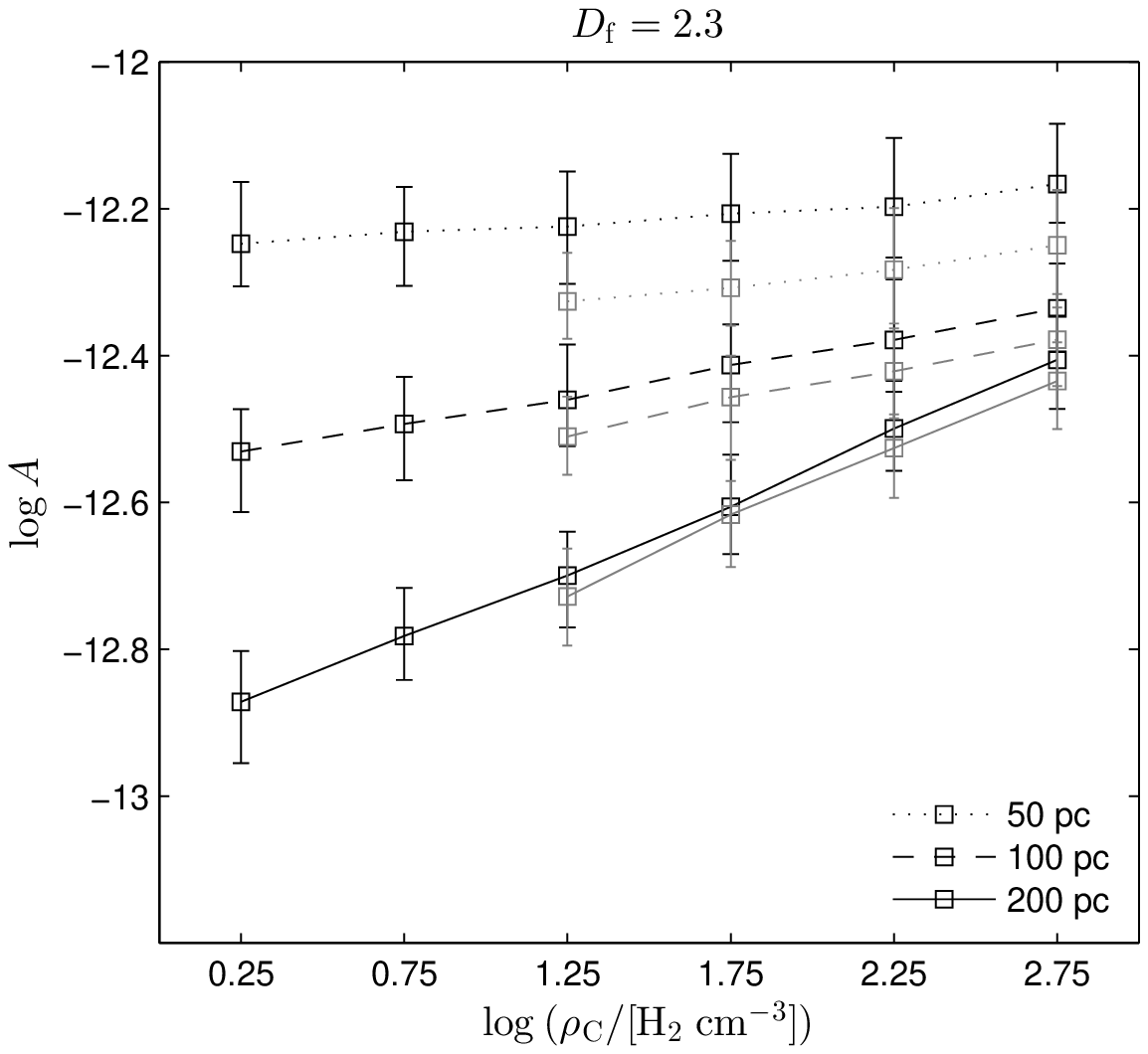}
	\includegraphics[width=0.48\textwidth]{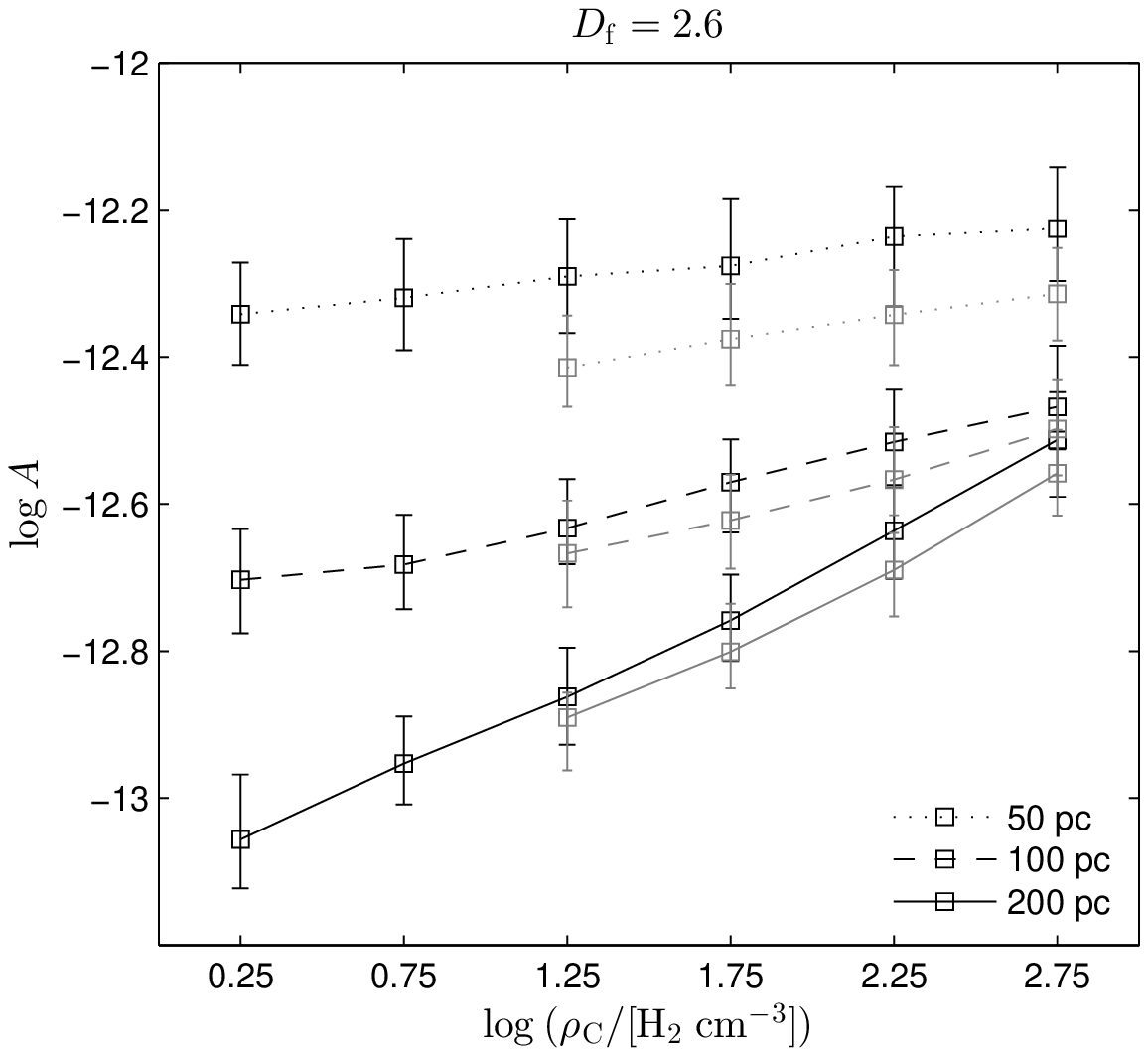}
	\caption{Model results. Slope $\beta $ (top graphs) and intercept $\log{A}$ (bottom graphs) vs. central density $\rho _{\rm{C}}$ (in logarithmic scale). Left graphs show results for the clouds of fractal dimension $D_{\rm{f}}=2.3$, while the right graphs are for $D_{\rm{f}}=2.6$. Dotted, dashed and solid line correspond to box edges of 50, 100 and 200 pc, respectively. Gray marks show the results for the case of having low density limit in the cloud, $\rho_{\rm{ICM}} = 10\;{{\rm{H}}_2}\,{\rm{cm}}^{-3}$.}
	\label{result}
\end{figure*}

The results of our model are graphically presented on Figure \ref{result}. The top two panels show the dependence of $\beta$ on the range of simulated central densities, while the bottom two show the same dependence for the intercepts $\log{A}$. Left side show results for the fractal dimension $D_{\rm{f}}=2.3$, right side for $D_{\rm{f}}=2.6$. For any of the top two graphs we see that this dependence looks linear and is more sensitive for larger cloud size. For smaller cloud sizes the dependence is weaker, because SNRs are more likely to have been expanded outside of the cloud into a very low or a ''zero`` density surrounding space. As we can clearly see, the remnants that formed in denser clumps of the cloud have steeper $\Sigma-D$ relation because the average density in the sphere encompassed by the remnant decreases faster with the diameter of the remnant. This means that initial ambient density (or, let's say, initial location in the cloud) plays important role in shaping the $\Sigma-D$ slope. It can also imply that the fractal nature of the ISM (inhomogeneity, to be more precise) can be partly responsible for scatter seen on $\Sigma-D$ diagrams, among the already mentioned causes like different (homogeneous) ISM density, energy of explosion and mass ejecta. This dependence on the initial ambient density, as the initial condition, is more likely to affect smaller SNRs. Large SNRs are less susceptible to density fluctuations because local density fluctuations can contribute far less to the value of the average density inside the SNR on large spatial scales. However, there is another relevant geometrical effect that has a particular importance for large SNRs, especially in smaller clouds. If a remnant forms near the edge of the cloud, part of it will eventually expand into a low density region. This will significantly reduce the average density inside the remnant which results in a steeper relation between this density and radius of the remnant. This affects the $\Sigma-D$ slope in the same way. It can be seen on the Figure \ref{result} as a dependence on the cloud size. For 50 pc clouds it is the dominant cause of high $\beta$ because all large remnants outsize the cloud. As we can also see from the results, values of $\beta$ are higher for $D_{\rm{f}}=2.6$ than for $D_{\rm{f}}=2.3$. According to \cite{Federrathetal2009}, this indicates that compressive (curl-free) driving of turbulence has stronger impact on the $\Sigma-D$ slope than solenoidal (divergence-free) driving. 
	
The gray marks on the plots show the results for the case of having a non-zero density in voids between the clumps ($\rho_{\rm{ICM}} = 10\;{{\rm{H}}_2}\,{\rm{cm}}^{-3}$), in accordance with the values from the literature \citep[see, for example,][]{Chevalier1999}. Considering that such an intervention violates the fractal structure and density contrasts of the cloud, we doubt that this may be accepted as an improvement of the model. Anyway, the results show that there is no substantial difference between these two models.

\section{Discussion}
\label{Discussion}
\subsection{The model}

Our simple geometric method for calculating the surface brightness does not include interactions between supernova remnants and inhomogeneous medium. Such a detailed model is beyond the scope of this paper and will be developed in the future, more elaborate work, along the lines of studies of \cite{BlandfordCowie1982,Chevalier1999,Bykov2000}. The way we calculated the surface brightness, by averaging the density inside the sphere, gives the same value of $\Sigma$ that the remnant of the same size would have in a homogeneous medium of the same density. However, the complexity of SNR evolution does not allow us to equate these two cases, especially since we still don't understand how the inhomogeneous medium affects the evolution. Such different environments would possibly affect the evolution of both diameter and surface brightness of an SNR differently. What can probably be assumed though is that the fractal environment of every remnant repeats the similar ''pattern`` of inhomogeneity, so that the negative effects of this approximation (for the quality of the results) probably cancel each other out to some extent.

\begin{figure*}
	\centering
	\includegraphics[width=0.48\textwidth]{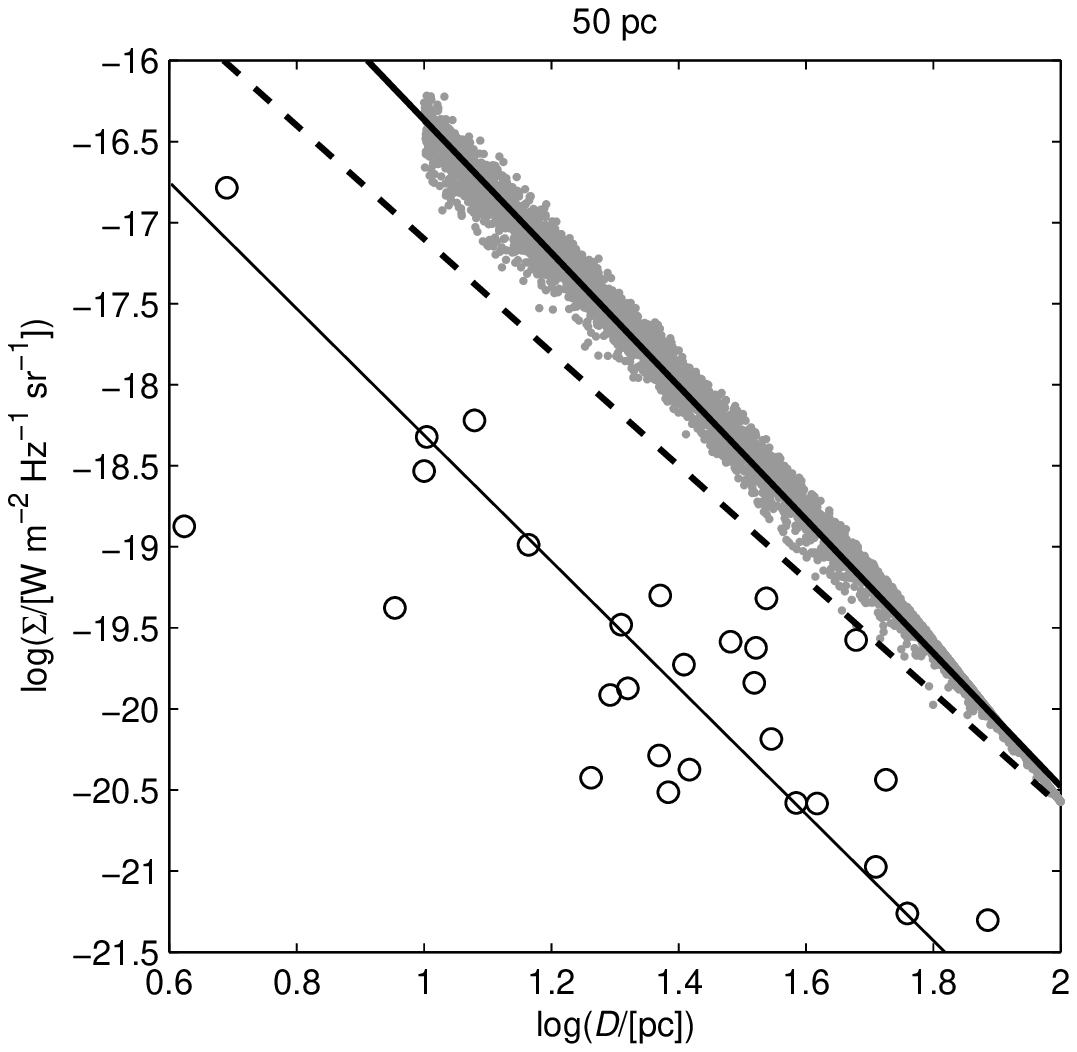}
	\includegraphics[width=0.48\textwidth]{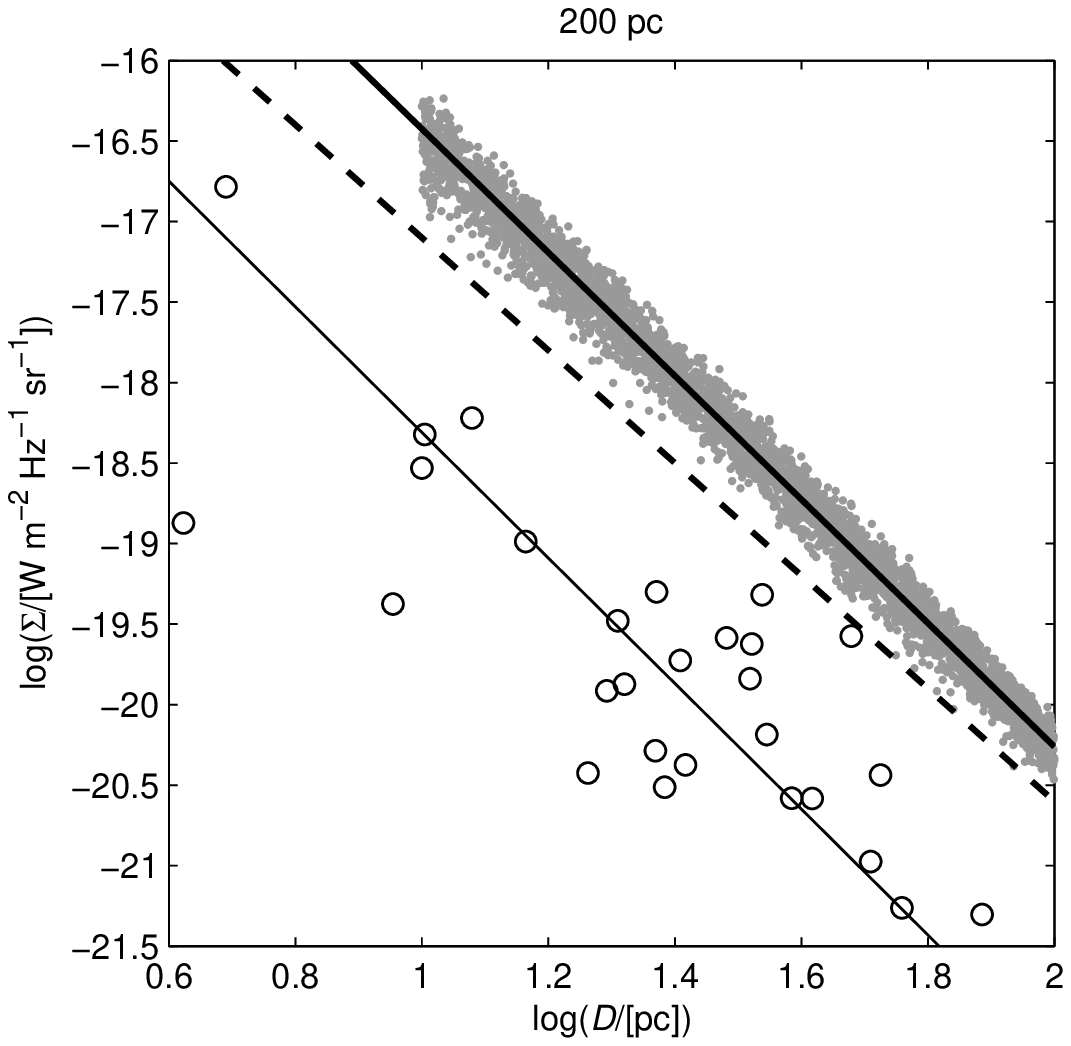} \\
	\includegraphics[width=0.48\textwidth]{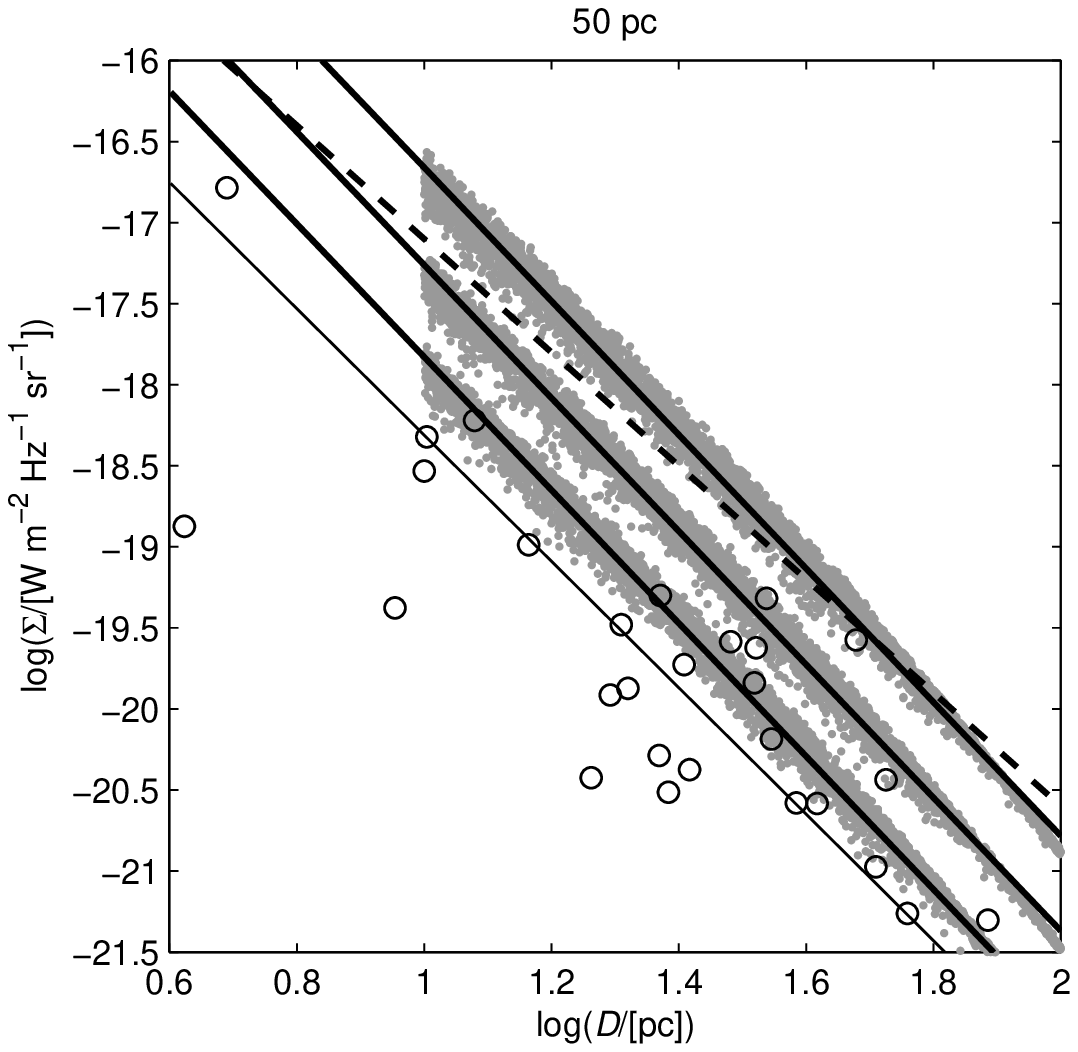}
	\includegraphics[width=0.48\textwidth]{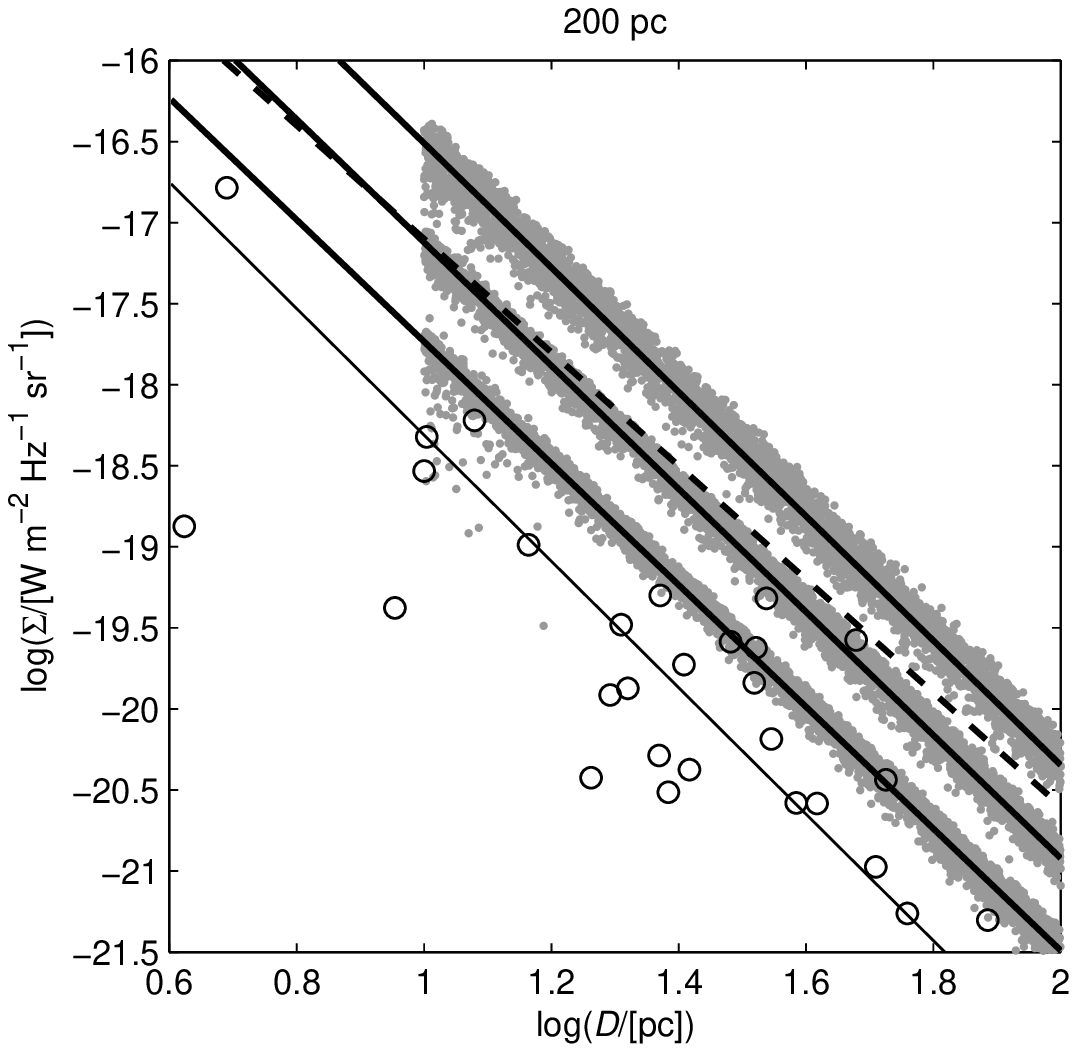}
	\caption{These $\Sigma-D$ plots show comparison between our data (for $D_{\rm{f}}=2.3$) and observational data. Our data are plotted as gray area of 4000 points with orthogonal fit (thick solid line). The thick dashed line represents the upper envelope from \citet{Berkhuijsen1986} (Eq. \ref{Berkhuijsen}). The circles and the thin solid line are observational data of SNRs interacting with molecular clouds from \citet{Pavlovicetal2013} and their $\Sigma-D$ fit (Eq. \ref{Pavlovic}). Top panels show data obtained when all density inside the sphere is included in calculation of $\Sigma$, while bottom panels show the data which only include densities below the thresholds $\rho_{\rm{th}}=10$, $100$ and $1000\;{{\rm{H}}_2}\,{\rm{cm}}^{-3}$ (from lower to upper fit, respectively). Left panels are for 50 pc cloud, right panels are for 200 pc cloud. The central densities range was $1-1000$ ${{\rm{H}}_2}\,{\rm{cm}}^{-3}$.}
	\label{fits}
\end{figure*}

Typical observational and theoretical values of coefficient $A$ are $\sim 10^{-17}-10^{-14}$. If we look at the obtained intercepts $\log{A}$ (Fig. \ref{result}), we see that they have unusually high values ($-13 < \log{A} < -12$). One reason is the steeper slope, but the main reason is that we obtain higher values for $\Sigma$ than normal. On Figure \ref{fits} we see an obvious discrepancy between the simulated and the observed $\Sigma$, by almost 2 orders of magnitude. This is because of a very high ambient density included in the calculation of $\Sigma$ (see Eq. \ref{sig}), several orders of magnitude above the values that are usually taken in the theoretical derivations of $\Sigma-D$ relation \citep[e.g. $\rho_0 = 10^{-24}$ g cm$^{-3} \approx 0.3\;{{\rm{H}}_2}\,{\rm{cm}}^{-3}$,][]{DuricSikvist1986}. The average density of 200 pc cloud in our model is $\approx 40\;{{\rm{H}}_2}\,{\rm{cm}}^{-3}$. The mass of a clump around one point (of a 3D Gaussian density structure) is $\approx 400\,M_\odot$ with the number density of $\approx 500\;{{\rm{H}}_2}\,{\rm{cm}}^{-3}$ in its centre, but it can be more than 10 times higher due to grouping of clumps. These values correspond to a typical densities in giant molecular clouds \citep{BertoldiMcKee1996,Chevalier1999}. However, such high values of surface brightness that we obtain in our simulations are not observed. \citet{Berkhuijsen1986} gave an upper envelope for the radio $\Sigma-D$ diagram such that for a given value of $\Sigma$, there is a maximal observable  diameter of an object (at the frequency $\nu=1\;{\rm{GHz}}$)
\begin{equation}
	{\Sigma _{1\;{\rm{GHz}}}} = 2.51 \times {10^{ - 14}}{D^{ - 3.5}}\;{\rm{W}}\,{\rm{H}}{{\rm{z}}^{ - 1}}\,{{\rm{m}}^{ - 2}}\,{\rm{s}}{{\rm{r}}^{ - 1}}. 
	\label{Berkhuijsen}
\end{equation}
It can be seen on Figure \ref{fits} (top graphs) that most of the points on our $\Sigma-D$ plots are above this line. The majority of the cloud mass resides in the cores of the clumps, which are not the sites of the significant radio synchrotron radiation, especially in the slower shocks of the older remnants. The clumps can be roughly described as having cold (dense) cores and warm coronae (warm neutral and warm ionized envelopes). The cores cannot be crushed when swept by the shock, so they stay embedded in the hot gas inside the remnant, while their envelopes eventually ''evaporate`` \citep{McKeeOstriker1977}. \cite{Chevalier1999} in his model of a remnant-molecular cloud interaction assumed that as a first approximation, a presence of the dense clumps can be neglected and that SNR primarily evolves in the interclump medium of density $n_{\rm{H}} = 5-25\;{\rm{H}}\,{\rm{cm}}^{-3}$. Using this model \cite{Bykov2000} distinguished interclump shock emission from molecular clump shock emission and found that particles reach higher energies in the interclump shock, so this region is dominant source of $\gamma$-ray and radio synchrotron emission in SNR, while dense clumps are expected to be sources of hard X-rays. Since the shock can hardly produce any fraction of ultrarelativistic electrons inside the cores of molecular clumps, the synchrotron radiation originating from there is practically negligible. On the other hand, the much lower density envelopes of the clumps can be significantly ionized by the shock (and also through the ionization precursor of the shock), contributing to synchrotron and other non-thermal radiation of the SNR. 
	
Considering these findings we ran the same simulations without taking into account the densities above a given threshold ($\rho_{\rm{th}}=10,\;100$ and $1000\;{{\rm{H}}_2}\,{\rm{cm}}^{-3}$) for the calculation of the average density in the sphere. This resulted in decreased values of $\Sigma$ and $\log{A}$, but the slope $\beta$ remained practically unchanged. These $\Sigma-D$ plots are shown at the bottom panels of Figure \ref{fits}. In the case of $\rho_{\rm{th}} \leq 100\;{{\rm{H}}_2}\,{\rm{cm}}^{-3}$, the points are under the upper envelope of \cite{Berkhuijsen1986}. For the density threshold of $10\;{{\rm{H}}_2}\,{\rm{cm}}^{-3}$ \citep[which corresponds to $5-25\;{\rm{H}}\,{\rm{cm}}^{-3}$ interclump density from][]{Chevalier1999} the fit is by about a half an order of magnitude (in $\Sigma$) above the fit from \cite{Pavlovicetal2013} (Eq. \ref{Pavlovic}), having $\log{A} \sim -13.7$ for 50 pc, and $\log{A} \sim -14$ for 200 pc cloud. This might be a successful test for the above assumptions that clumps do not participate in producing radio synchrotron radiation. Even if we neglect the majority of the molecular cloud mass (residing in cores of the clumps), the surface regions of the clumps still have the fractal distribution throughout the volume (one can think of them as convoluted density isosurfaces). Consequently, the slope of the density-size relation stays the same, which reflects on the slope of the $\Sigma-D$ relation, too.

In our simulations, the progenitor stars are all identical (because we used one unique equation for the surface brightness) and we interpret them as massive stars according to their place of origin (SNR centres are in dense medium). In reality, however, we could not ignore the impact that such stars would have on their environment. For example, O stars have sufficient photoionizing radiation and wind power to clear out the surrounding molecular matter to distance of more than 15 parsecs \citep[up to $\sim 28\;{\rm{pc}}$,][]{Chevalier1999}, which would significantly change the $\Sigma-D$ plots in our simulation, at least for SNRs with smaller diameters. On the other hand, there are indications that remnants residing in the dense medium enter the radiative phase much earlier and are correspondingly smaller \citep{Asvarov2014,BandieraPetruk2010,Badenesetal2010,Vink2012,Slaneetal2015}, which could affect the $\Sigma-D$ relation in the domain of larger diameters.

\subsection{Geometrical aspects of the ISM model and its consequences for other $\Sigma-D$ relations}
\label{geometrija}
In this subsection we discuss the specific properties of fractal clouds and their impact on the $\Sigma-D$ relation of our model. The average density in the fractal cloud that we used decreases with size scale as
\begin{equation}
	\rho  \propto {S^{{D_{\rm{f}}} - 3}}
\end{equation}
which follows directly from the mass-size relation (\ref{M-S}). In other words, the average density of the (e.g. spherical) fragment within the cloud decreases as a power law (for $D_{\rm{f}}<3$) as the fragment radius increases (i.e. as the sphere grows towards the scale of the next level of hierarchy). This means that, for $D_{\rm{f}}=2.3$, the density inside the SNR spheres in our fractal cloud depends on the SNR diameter as
\begin{equation}
	\rho  \propto {D^{ - 0.7}}.  \label{ro-D}
\end{equation}
However, this relation applies only when the centre of the sphere coincides with some local peak of density and if the sphere does not extend outside the cloud. For these reasons we should ignore results for the smaller clouds and focus only on the 200 pc cloud. As the distance between the local peak and the sphere centre increases (which is accompanied by the decrease of $\rho _{\rm{C}}$) the deviation from Eq. (\ref{ro-D}) is larger. We can rewrite it as
\begin{equation}
	\rho  \propto {D^{ - 0.7 + \delta }}. \label{delta}
\end{equation}
The exponent variation, $\delta  = \delta \left( {{\rho _{\rm{C}}}} \right)$ (which is around zero for regions where $\rho_{\rm{C}}$ is close to local maximum in density field and increases for lower values), affects the slope $\beta$ of our model. When relation (\ref{delta}) is included in theoretical $\Sigma-D$ relation (\ref{sig}) we get:
\begin{equation}
	\Sigma \left( D \right) \propto {\rho ^{0.5}}{D^{ - 3.5}} \propto {\left( {{D^{ - 0.7 + \delta }}} \right)^{0.5}}{D^{ - 3.5}}. 
\end{equation}
Thus, we obtain
\begin{equation}
	\Sigma \left( D \right) \propto {D^{ - 3.85 + {\delta _1}}}, \left( {{\delta _1} = 0.5\,\delta } \right).
\end{equation}
This means that, as $\delta \ge 0$, the maximum value of $\beta$ should be 3.85. The fact that $\beta$ is higher than this for some values of $\rho_{\rm{C}}$ as seen on Figure \ref{result} (top-left graph), is probably due to overlapping of some clumps in the cloud (these variations might be due to random placement of points in formation of a fractal and selection of a smoothing kernel from Equation \ref{polje}) as well as due to already explained effect of SNR expanding outside of the cloud (it happens in 200 pc clouds as well).

From this we see that whatever theoretical $\Sigma-D$ relation of the form $\Sigma \left( {\rho ,D} \right) \propto {\rho ^\eta }{D^{ - {\beta _0}}}$ is used, our simulations would give 
\begin{equation}
\Sigma \left( D \right) \propto {\rho ^\eta }{D^{ - {\beta _0}}} \propto {D^{ - 0.7\eta  - {\beta _0} + \eta \delta }},
\end{equation}
with the resulting slope
\begin{equation}
\beta  = {\beta _0} + \eta \left( {0.7 - \delta } \right),
\end{equation}
or, more generally,
\begin{equation}
\beta  = {\beta _0} + \eta \left( {3 - D_{\rm{f}} - \delta } \right). \label{beta-Df}
\end{equation}
Finally, this equation clearly shows the connection between the $\Sigma-D$ slope and the fractal dimension of the environment, and explains the discrepancy of the results for different values of fractal dimension. Of course, our results do not show the exact difference of $0.5(2.6-2.3)=0.15$ between the two cases probably because of the influence of other effects, especially the possible dependence $\delta(D_{\rm{f}})$ and the different average distance of the same central densities from the density peaks.

The selected $\Sigma-D$ relation from \citet{DuricSikvist1986} is suitable for this kind of testing, i.e. it depends on the square root of ambient density and its $\beta$ doesn't vary extremely from the observed values \citep[see empirical relation from][]{Pavlovicetal2013}. Besides, it is parametrized with one set of $A$ and $\beta$ values for the whole domain of our interest. \citet{BerezkoVolk2004} divided the Sedov phase into \textit{early} and \textit{late} subphases that mutually overlap without a clear distinction between them. In the early Sedov phase, $\Sigma$ practically does not depend on density while the late Sedov phase is characterized by $\Sigma  \propto {\rho ^{0.75}}{D^{ - 2}}$, which would in our model result in a relation $\Sigma  \propto {D^{ - 2.525 + 0.75{\delta}}}$. 

\begin{figure}
	\centering
	\includegraphics[width=0.45\textwidth]{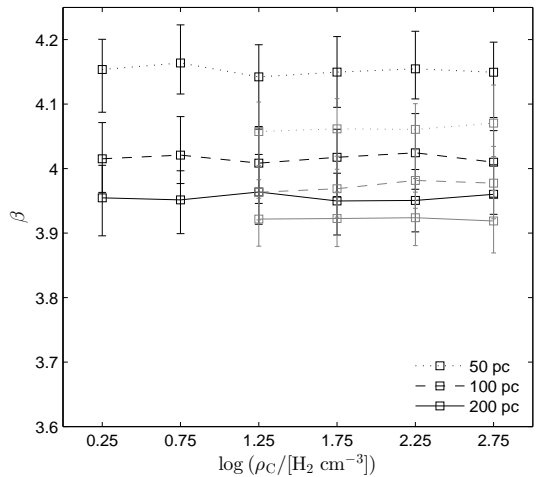}
	\caption{Model results for $\sigma  = 0.001{S_{{\rm{cl}}}}$. Figure legend is the same as in Figure \ref{result}. ($D_{\rm{f}}=2.3$)} \label{sig0001}
\end{figure}

Also, our results are significantly dependent  on the adopted value of the smoothing parameter $\sigma$ in the Eq. (\ref{polje}). The convolution of fractaly distributed points within the box with Gaussian kernel gives a continuous density distribution throughout the whole volume of the box. The value of $\sigma$ must not be too big to unnecessarily homogenize the fractal and thus change its properties, but it should be neither too small to make the convolution meaningless. While ''fine-tuning`` our model, we found that best results are achieved if the selected $\sigma$ value is approximately equal to half mean distance between neighbouring points in the last level of hierarchy and this value is also of the same order of magnitude as in \citet{Sanchez2006}. If $\sigma$ was too small, all SNR centres would fall very close to the local peaks in density, so we would have $\delta \to 0$ for any $\rho_{\rm{C}}$, which happens at $\sigma \approx 0.001S_{\rm{cl}}$ (see Figure \ref{sig0001}). This means that in such case there would be no dependence of $\beta $ on the central density ${\rho _{\rm{C}}}$.

\begin{figure}
	\begin{center}
		\includegraphics[width=0.45\textwidth]{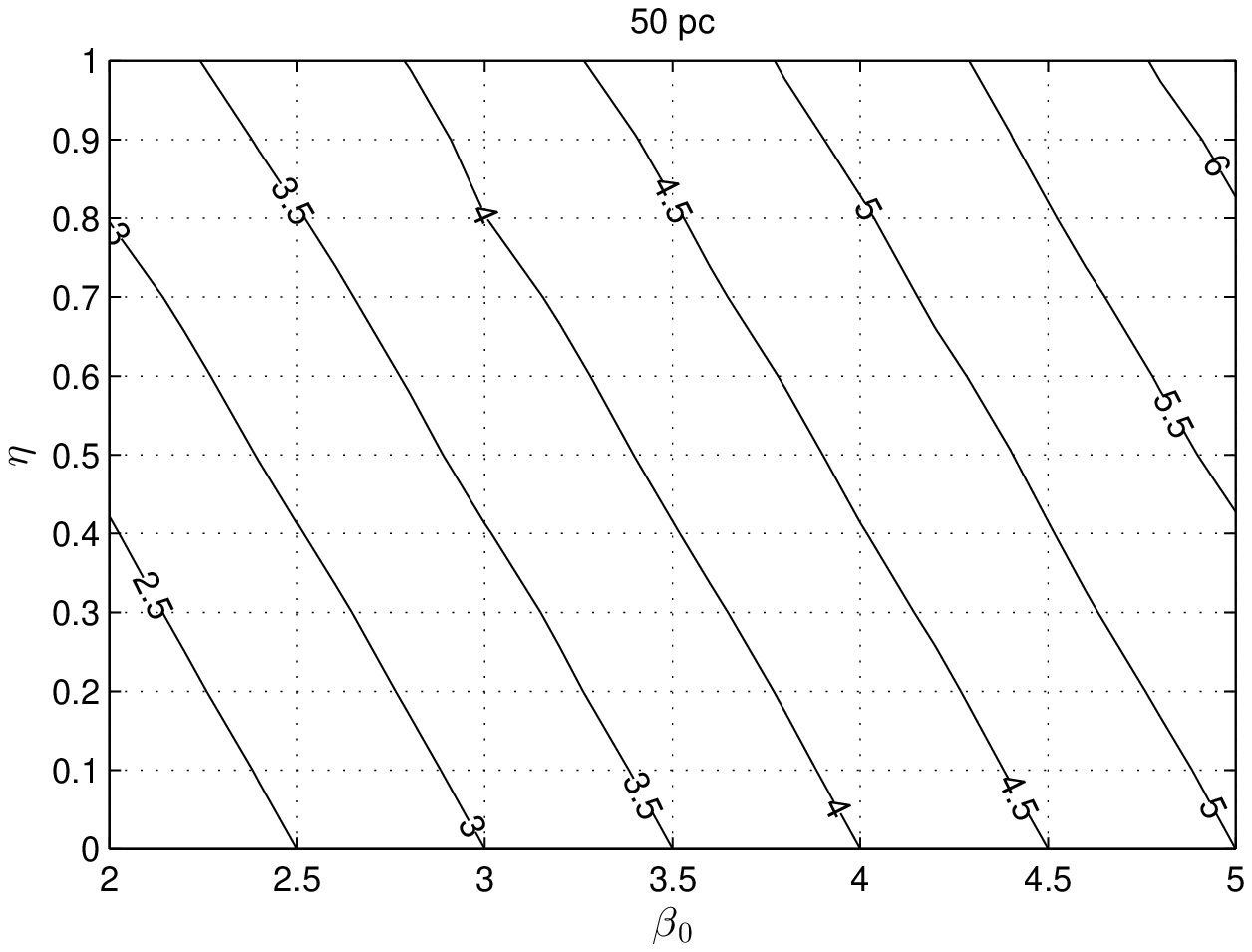}
		\includegraphics[width=0.45\textwidth]{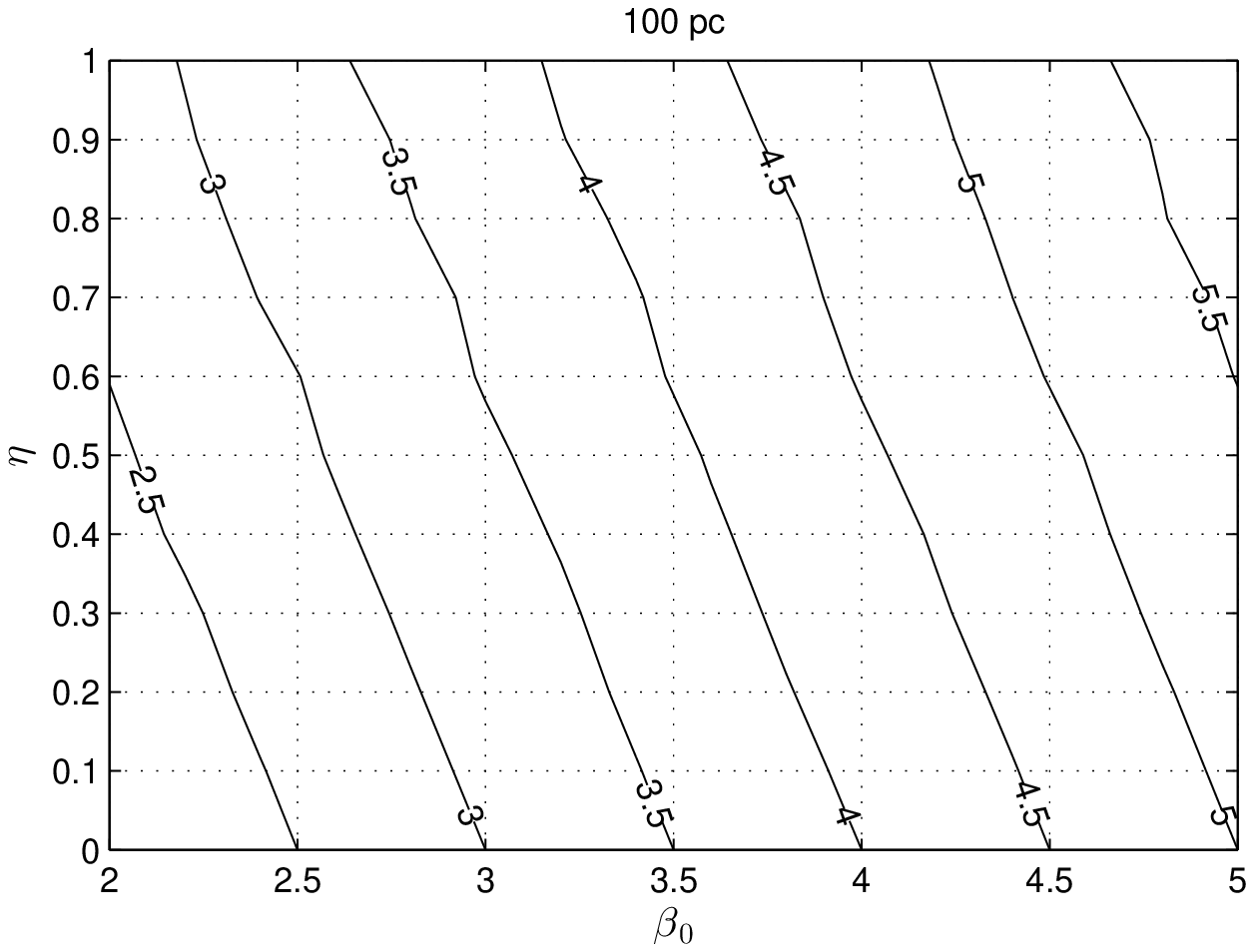}
		\includegraphics[width=0.45\textwidth]{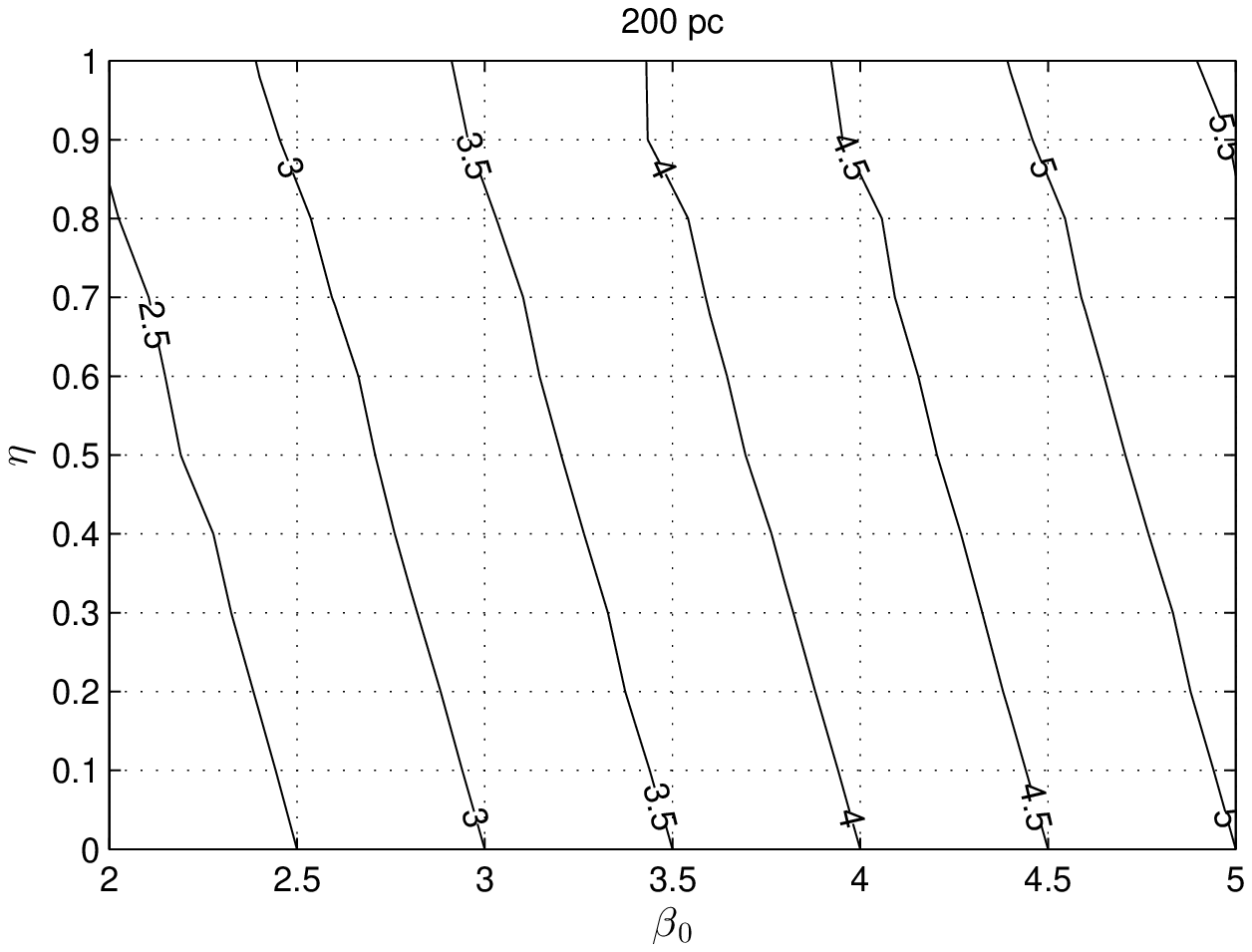}
		\caption{The contour plots of obtained slopes $\beta$ for corresponding $\eta$ and $\beta_0$ of general theoretical $\Sigma \left( {\rho ,D} \right) \propto {\rho ^\eta }{D^{ - {\beta _0}}}$ relation in the given range. The central densities range was $1-1000$ ${{\rm{H}}_2}\,{\rm{cm}}^{-3}$. Each $\beta$ was obtained as a median from 30 simulations. From top to bottom, graphs are for 50, 100 and 200 pc clouds. ($D_{\rm{f}}=2.3$)} \label{kont}
	\end{center}
\end{figure}

In order to get a better insight on how the fractal medium affects the $\Sigma-D$ relation in general we make a contour plot in Figure \ref{kont}. Contours of $\beta$ obtained from our simulations are plotted in the parameter space of $\eta$ and $\beta_0$ from  $\Sigma \left( {\rho ,D} \right) \propto {\rho ^\eta }{D^{ - {\beta _0}}}$ general form of the theoretical relation. On these plots it can be seen that the sensitivity of $\beta$ on $\eta$ depends on the cloud size. This means that for the same $\eta$, the difference between $\beta$ and $\beta_0$ is higher for smaller clouds. Thus, if the size of the cloud is substantially smaller than the remnant (top plot), $\beta-\beta_0$ strongly depends on $\eta$. On the other hand, if the size of the remnant is substantially larger than the cloud (bottom plot), this dependence is weaker but still exists. We can interpret this as young (smaller than cloud) and old (larger than cloud) remnants and conclude that ambient density has more significant influence on the slope of the $\Sigma-D$ relation for older SNRs originating from molecular clouds (with a prior condition of any $\Sigma$ dependence on ambient density). This conclusion can also be seen clearly on $\Sigma-D$ plots for 50 pc cloud on Figure \ref{fits}. However, besides these fine differences, all three panels from Figure \ref{kont} imply that, with regard to the $\Sigma$ dependence on ambient density, the empirical $\Sigma-D$ slopes can be expected to be steeper than the slopes of their theoretical counterparts.  

\section{Conclusions}
The aim of this study was to show that the slope of $\Sigma-D$ relation is affected by variations in density of the environment in which the remnant evolves. This dependence is given on Figure \ref{result}. A significant result is that this influence is qualitatively demonstrated. Within the assumptions of the model, it is evident that some empirical slopes that are steeper than theoretical can be partly explained by the ambient density fractal structure for supernova remnants expanding in dense ISM areas of molecular clouds. 

Assuming that the fractal model gives a fairly realistic density structure of molecular clouds, we showed that the slope $\beta$ of $\Sigma-D$ relation correlates (steepens) with the density increase of the region where the remnants have formed. A simple effects of geometry of a fractal structure of ISM can contribute (with respect to $\Sigma$ dependence on ambient density) to a significantly steeper slopes  of empirical $\Sigma-D$ relations. This is more pronounced for larger (older) remnants but can be significant even when smaller (younger) remnants are considered. 

We have also touched on the subject of driving modes of the interstellar turbulence. As the compressive driving forms the structures with lower fractal dimension \citep{Federrathetal2009}, it is likely that this mode of driving affects the $\Sigma-D$ slope more than the solenoidal driving mode. This can be seen from the results (Figure \ref{result}) as well as from the derived relation between $\beta$ and $D_{\rm{f}}$ (see Equation \ref{beta-Df}).

More detailed model should include the impact that progenitor stars would have on ambient medium as well as detailed consideration of radiation production by shock waves within the inhomogeneous medium.

\section*{Acknowledgements}
We thank an anonymous referee for insightful comments that have substantially improved the quality of the paper. BV, DU, BA and TP acknowledge financial support from the Ministry of Education, Science and Technological Development of the Republic of Serbia through the project \#176005 ''Emission nebulae: structure and evolution``.




\bibliographystyle{mnras}
\bibliography{mybib} 



\bsp	
\label{lastpage}
\end{document}